\documentclass[mathleft
]{an}
\usepackage{graphicx}
\usepackage{times}
\begin{document}

\Pagespan{001}{}
\Yearpublication{2010}%
\Yearsubmission{2010}%
\Month{09}%
\Volume{999}%
\Issue{00}%

\title{Hard and soft spectral states of ULXs}


\author{R. Soria\inst{1}\fnmsep\thanks{Corresponding author:
  \email{roberto.soria@mssl.ucl.ac.uk}\newline}
}
\titlerunning{Hard and soft spectral states of ULXs}
\authorrunning{R. Soria}
\institute{
MSSL, University College London, Holmbury St Mary, Surrey, RH5 6NT, UK}

\received{28 Sep 2010}
\accepted{}
\publonline{later}

\keywords{black hole physics -- accretion, accretion disks -- X-ray:binaries}

\abstract{%
I discuss some differences between the observed spectral states of ultraluminous X-ray sources (ULXs) 
and the canonical scheme of spectral states defined in Galactic black holes.
The standard interpretation of ULXs with a curved spectrum, or a moderately steep 
power-law with soft excess and high-energy downturn, is that they are an extension 
of the very high state, up to luminosities $\approx 1$--$3 L_{\rm Edd}$. 
Two competing models are Comptonization in a warm corona, and slim disk; 
I suggest bulk motion Comptonization in the radiatively-driven outflow 
as another possibility. The interpretation of ULXs with a hard power-law spectrum 
is more problematic. Some of them remain in that state over a large range 
of luminosities; others switch directly to a curved state without going 
through a canonical high/soft state. I suggest that those ULXs are 
in a high/hard state not seen in Galactic black holes; that state may overlap 
with the low/hard state at lower accretion rates, and extend 
all the way to Eddington accretion rates. If some black holes can reach 
Eddington accretion rates without switching to a standard-disk-dominated state, 
it is also possible that they never quench their steady jets.} 

\maketitle

\section{Introduction}

The accretion/outflow activity of black holes (BHs) 
is usually classified into states, using phenomenological 
and/or physical criteria. In particular, Galactic BH X-ray binaries 
have been used to define a set of ``canonical'' states (Remillard \& McClintock 2006; 
Belloni 2010a; Done, Gierlinski \& Kubota 2007; Homan \& Belloni 2005; 
Fender, Belloni \& Gallo 2004).
Based on their X-ray appearance, BH spectra in the low/hard state 
are dominated by a hard power-law, 
sometimes with a weak contribution from a cool and/or truncated disk; 
in the high/soft state, they are dominated by an optically thick, 
geometrically thin accretion disk, sometimes with a weak 
contribution from a steep power-law; in the very high state, 
the steep power-law becomes dominant over the disk component.
Based on their power output, BHs can be in  
a radiatively inefficient state dominated by advection or jets 
at lower accretion rates (corresponding to the low/hard state); 
a radiatively efficient state at higher accretion rates 
(corresponding to the high/soft state); and again in a less and less 
efficient state as the accretion rate approaches or exceeds the Eddington 
limit. 

The physical structure of the high/soft state is comparatively well known, 
and tight relations link the accretion rate and inner disk radius 
(which constrains the BH mass) to observable quantities such as disk luminosity 
and maximum temperature (Makishima et al.~2000).
The physical structure of the low/hard and very high states is less clear; 
they may in fact be ill-defined labels corresponding to a variety of scenarios.
For example, the low/hard state has been interpreted either as due 
to an advection-dominated accretion flow (ADAF) plus truncated disk, 
or to a hot, optically-thin Comptonizing corona above a cool disk, 
or to a synchrotron-emitting jet, or to hot shocked gas 
at the centrifugal boundary layer ({\it e.g.}, respectively, 
Esin, McClintock \& Narayan 1997; Liang \& Price 1977; Markoff et al.~2003; 
Markoff, Nowak \& Wilms 2005; Chakrabarti \& Titarchuk 1995).
Similarly, the very high state has been interpreted in different ways: 
either due to Comptonization in a marginally optically-thick corona 
($\tau \sim 1$) covering the inner disk (Done \& Kubota 2006); 
or to the inner disk itself becoming effectively-optically-thin 
and scattering-dominated (Artemova et al. 2006); or to a combination 
of radiation trapping, advection and outflows at near-Eddington 
accretion rates (slim disk models: 
Watarai, Mizuno \& Mineshige 2001; Ohsuga et al.~2009).

Despite the physical uncertainties, as a first-order approximation 
there is general agreement that: 
a) when the mass accretion rate increases above 
$0.03$--$0.1$ Eddington, most Galactic BHs switch from the hard 
(power-law dominated) to the soft (thermal disk dominated) state, 
and spend the majority of their outburst phase in the latter state; 
b) the jet is suppressed during this transition, as the inflow collapses 
into a geometrically thin disk without a poloidal magnetic field component 
(Fender et al.~2004; Belloni et al.~2005).

\section{ULX spectral state(s)}

X-ray spectra of ultraluminous X-ray sources (ULXs) have been studied 
so far almost only in the $0.3$--$10$ keV band, because of observational 
limitations (lack of sensitivity and spatial resolution at higher energies);
multiband information (optical/UV, radio) is also very sketchy.
Their classification into canonical states is much more uncertain 
than for Galactic BHs. We can summarize them with the following scheme 
(Makishima 2007): \\
a) simple power-law, with a hard photon index $\Gamma \approx 1.0$--$2.0$, 
similar to the low/hard state of Galactic BHs, but with 
a broader distribution of photon indices;\\
b) a power-law-like shape with a slope $\Gamma \approx 2.0$--$2.7$ 
in the $\approx 1$--$5$ keV range, a steepening or downturn 
at energies $\ga 5$ keV, and a soft excess below 1 keV (typically modelled 
with a thermal component with $kT \approx 0.1$--$0.2$ keV). 
This is perhaps the most common state for luminous ULXs, and the most 
distinctive with respect to Galactic BHs; that is why it is sometimes 
referred to as the ``ultraluminous state'' (Gladstone, Roberts \& Done 2009; 
Walton et al.~2010);\\
c) a broad, slightly ``curved'' state, not well fitted by a single 
power law but open to a variety of other fitting models. In 
many cases, a Comptonization model with a low-temperature corona 
($kT_e \approx 2$ keV) or a non-standard disk model 
({\it e.g.}, slim disk) offer equally acceptable fits 
(Roberts 2007; Stobbart, Roberts \& Wilms 2006). In other cases, 
the data favour the Comptonization model (Gladstone at al.~2009). 
And in some cases, disk reflection has been suggested as an alternative 
explanation for the curvature (Caballero-Garcia \& Fabian 2010; Walton et al.~2010);\\
d) a supersoft thermal spectrum, with $kT_{\rm bb} \approx 70$--$100$ eV 
and most of the luminosity emitted in the far UV rather than X-rays.
This is seen only in a handful of transient sources (Soria \& Ghosh 2009, 
and references therein), which may represent an altogether different population. 
For example, they might be short-lived super-Eddington 
outflow phases of stellar-mass BHs (Carpano et al.~2007). I am not going 
to consider this state further in this paper.

The first thing to notice is that the difference between 
the first three states (in particular, b and c) is blurred. 
At lower signal-to-noise ratios, all spectra are power laws.
At slightly higher signal-to-noise ratios, a broad curvature can easily 
be modelled also with a broken power-law or a power-law with 
an exponential cutoff. Moreover, ULXs seem to have a more continuous 
distribution of features, rather than being grouped into really 
separate states.
There are at least three observational differences 
between the X-ray spectral classifications of ULXs 
and Galactic stellar-mass BHs:\\
i) several ULXs are found in a hard power-law-dominated 
state, at X-ray luminosities $\approx 10^{40}$ erg s$^{-1}$ (Winter, Mushotzky 
\& Reynolds 2006; Berghea et al.~2008). Some of them 
may vary by a factor of a few over months/years without spectral changes 
(Soria et al.~2009; Kaaret \& Feng 2009); 
others may show a change in slope and curvature (as discussed in Section 4). 
I speculate that those power-law ULXs are in a steady high/hard state, 
unknown in Galactic BHs. It is not known whether a steady, powerful jet 
is associated to this luminous hard-power-law state;\\
ii) no ULXs (except probably for M\,82 X41.4$+$60: Feng \& Kaaret 2010) 
have been found in a canonical high/soft state: that is, even when 
their spectra are formally fitted with a diskbb model, 
their temperature-luminosity relations are inconsistent 
with the requirements of a standard disk (typically, 
their temperatures are a factor of 2 or 3 too high).
Perhaps their accretion rate is always too high 
to allow for the presence of a standard disk; however, 
some ULXs have occasionally declined in luminosity 
by an order of magnitude, and re-brightened, without 
being seen to spend any time in the high/soft state;\\
iii) many ULXs with a spectral curvature show a characteristic 
downturn at energies $\approx 5$--$6$ keV. This can be interpreted 
as a characteristic colour temperature $\approx 1.5$--$3$ keV in the innermost 
part of the accretion disk, or as a characteristic electron 
temperature $\approx 1.5$--$3$ keV in the Comptonizing region 
(warm, optically-thick corona: Stobbart et al.~2006; Gladstone et al.~2009).
Either way, this characteristic temperature is unknown 
in Galactic BHs, where disk temperatures are $\la 1$ keV and 
coronal temperatures are $\sim 100$ keV. 

In the rest of this paper, I will briefly discuss some implications 
of these differences.

\section{ULXs with curved spectra}

The general interpretation of ULXs with a steeper spectrum, soft excess, 
various degrees of spectral curvature and downturn above 5 keV 
({\it i.e.}, states b and c) is converging 
to a consensus, even though the details of the model for each source may differ.
They appear to be accreting at levels above the upper limit for 
a standard disk, near or up to a few times the Eddington luminosity.
Their state can be described either as an extension of the very high state, 
with a cooler but optically thicker corona replacing/covering the inner disk 
(Done \& Kubota 2006; Gladstone et al.~2009); 
or as a modification of the inner disk itself, now dominated by radiation pressure, 
electron scattering, energy advection through radiation trapping, 
outflows (slim disk model: Watarai et al.~2001; 
supercritical accretion model: Poutanen et al.~2007).
 
ULXs in these kinds of states have characteristic luminosities 
from $\approx$ a few $10^{39}$ to $\approx 2 \times 10^{40}$ erg s$^{-1}$, and are never 
seen in a canonical high/soft state. In fact, as we mentioned above, 
there are no known high/soft sources above $\approx 5 \times 10^{39}$ erg s$^{-1}$, 
except perhaps for M82 X41.4$+$60. This suggests that the ULX population contains 
very few or no accreting BHs with masses 
$\ga 100 M_{\odot}$: if there were many of them, some would likely have been seen 
in their high/soft state in that luminosity range. On the other hand, 
if ULXs had masses $\la 30 M_{\odot}$, 
the accretion rate required to reach luminosities $\sim 10^{40}$ erg s$^{-1}$ 
would be implausibly high, because $L \sim L_{\rm Edd} (1 + 0.6 \ln \dot{m})$ 
(Shakura \& Sunyaev 1973; Poutanen et al.~2007).

The typical flux contribution of the soft excess, interpreted as the emission 
from the standard outer disk, is $\sim 10\%$ (order-of-magnitude) of the total 
X-ray flux, suggesting that $\sim 90\%$ of the flux comes 
from the non-standard inner region, and hence, that the transition radius 
where the disk becomes non-standard, or covered/replaced by a scattering corona 
is $R \sim 10 R_{\rm isco}$ (Soria 2007). This is self-consistent with the observed characteristic 
temperature of the soft excess, $kT_{\rm se} \sim 0.1$--$0.2$ keV. 
A transition radius $R \sim 10 R_{\rm isco}$ also suggests a characteristic 
accretion rate $\dot{m} \sim 10$ in Eddington units, which implies 
characteristic luminosities $\sim 2$--$3 L_{\rm Edd}$, and a characteristic mass  
$\sim 50 M_{\odot}$ (within a factor of 2), in agreement 
with stellar-evolution arguments.

The characteristic temperatures of the spectral downturn 
is consistent with both the warm-corona thermal Comptonization scenario
and with the slim disk scenario. Here I suggest another alternative scenario. 
I speculate that the curved spectra may be the signature  
of bulk motion Comptonization (Laurent \& Titarchuk 1999,2007) 
in a radiatively driven outflow, 
which is to be expected at super-Eddington accretion rates. 
If so, we could interpret the ``corona'' temperature 
as the characteristic kinetic energy of the outflow, 
$(1/2) m_e v^2 \approx (3/2)kT_e \approx 4$ keV, 
which implies an outflow speed $v \approx 40,000$ km s$^{-1}$. That is 
a plausible value for the outflow velocity: as a comparison, 
the escape velocity from $R \sim 10 R_{\rm isco}$ is 
$\approx 50,000$ km s$^{-1}$ for a Schwarzschild BH.
That is indeed the region where we expect the disk to puff up 
and an outflow to be launched, when $\dot{m} \sim 10$.

\section{ULXs in a hard state}

Several ULXs are seen in a hard power-law-dominated state at high luminosities; 
a selection is listed in Table 1 (see Feng \& Soria 2011, 
in prep., for a more comprehensive list). Their interpretation is more problematic, 
because hard power laws are not known to be associated with very high states 
of Galactic BHs.

\begin{table}
\begin{center}
\begin{tabular}{lccr}
\hline
Source & $\Gamma$   & $L_{0.3-10}$ & Reference\\
 &   & ($10^{40}$ erg s$^{-1}$) & \\
\hline\\[-5pt]
NGC\,5775 X1 & $1.8^{+0.3}_{-0.2}$ & $7.5$ & (1)\\[3pt]
NGC\,5775 X2 & $1.4^{+0.2}_{-0.1}$ & $1.6$ & (2)\\[3pt]
             & $1.5^{+0.3}_{-0.2}$ & $0.8$ & (2)\\[3pt]
Arp 240 X1   & $1.5^{+0.5}_{-0.5}$ & $7$ & (3)\\[3pt]
NGC\,1365 X1 & $1.7^{+0.1}_{-0.1}$ & $2.8$ & (4)\\[3pt]
             & $1.8^{+0.1}_{-0.1}$ & $0.5$ & (4)\\[3pt]
NGC\,1365 X2 & $1.2^{+0.3}_{-0.2}$ & $3.7$ & (4)\\[3pt]
             & $1.1^{+0.1}_{-0.1}$ & $0.15$ & (4)\\[3pt]
Antennae X11 & $1.8^{+0.1}_{-0.1}$ & $2.1$ & (5)\\[3pt]
             & $1.7^{+0.1}_{-0.1}$ & $1.4$ & (5)\\[3pt]
Antennae X16 & $1.4^{+0.1}_{-0.1}$ & $1.8$ & (5)\\[3pt]
             & $1.2^{+0.1}_{-0.1}$ & $0.7$ & (5)\\[3pt]
Antennae X42 & $1.7^{+0.1}_{-0.1}$ & $1.0$ & (5)\\[3pt]
Antennae X44 & $1.7^{+0.1}_{-0.1}$ & $1.3$ & (5)\\[3pt]
M82 X42.3+59 & $1.4^{+0.1}_{-0.1}$ & $1.5$ & (6)\\[3pt]
M99 X1       & $1.7^{+0.1}_{-0.1}$ & $1.8$ & (7)\\[3pt]
NGC 1068 X1  & $0.9^{+0.1}_{-0.1}$ & $1.5$ & (8)\\[3pt]
NGC 3628 X1  & $1.8^{+0.2}_{-0.2}$ & $1.2$ & (9)\\[3pt]
\hline
\end{tabular}
\end{center}
References: (1): Li et al.~(2008); (2): Ghosh et al.~(2009); (3): Swartz et al., in prep.; 
(4): Soria et al.~(2009); (5): Feng \& Kaaret (2009); (6): Feng, Rao \& Kaaret (2010); (7): Soria \& Wong (2006);
(8): Smith \& Wilson (2003); (9): Strickland et al.~(2001).\\[3pt]
\caption{Selection of ULXs in a hard power-law state; some have been observed in more than one epochs. 
I used WebPimms to convert all unabsorbed luminosities to the same energy band.}
\end{table}

If they were in a canonical low/hard state and had the same state 
transition properties as Galactic BHs,  
their Eddington luminosities would be $\ga 10^{42}$ erg s$^{-1}$ 
and their BH masses $\ga 10^4 M_{\odot}$ (Winter et al 2006). 
Furthermore, we would have to conclude that their accretion rate, 
although high, never reaches the hard/soft transition threshold.
However, there is still no observational evidence or compelling 
theoretical arguments for this range of BH masses in ULXs; 
here I take the conservative view that most ULXs are powered 
by BHs with masses $\la 100 M_{\odot}$ formed from local 
stellar-evolution processes. Also, there are no astrophysical 
examples of other populations of accretion-powered sources 
with an upper luminosity limit $\sim 0.1 L_{\rm Edd}$ 
rather than $L_{\rm Edd}$. 
Thus, placing the hard power-law state ``below'' the (unseen) 
canonical high/soft state presents serious difficulties.

A few of the hard power-law ULXs have been detected with a curved spectrum 
at other epochs. The best-known examples are IC342 X1 and X2, observed by {\it ASCA} 
in 1993 and 2000 (Kubota et al.~2001). X1 was in a curved/slim-disk state   
($kT_{\rm in} \approx 1.8$ keV) in 1993, with $L_{0.3-10} \approx 2 \times 10^{40}$ erg s$^{-1}$; 
it switched to a hard power-law state  ($\Gamma \approx 1.7$) in 2000, with 
$L_{0.3-10} \approx 6 \times 10^{39}$ erg s$^{-1}$. Conversely, 
X2 was in a hard power-law state ($\Gamma \approx 1.4$) in 1993, with 
$L_{0.3-10} \approx 8 \times 10^{39}$ erg s$^{-1}$; it switched to 
a curved/slim-disk state ($kT_{\rm in} \approx 1.6$ keV) in 2000, with 
$L_{0.3-10} \approx 1.5 \times 10^{40}$ erg s$^{-1}$. Further switches 
between a hard power-law state and a curved/slim-disk state were found 
in both sources, from a series of {\it XMM-Newton} and {\it Chandra} observations 
between 2001 and 2006 (Mak, Pun \& Kong 2010).
If the hard power-law state corresponded to the low/hard state, 
and the curved state to the very high state, the narrow range 
of luminosities over which the transition occurs would not leave 
much space for a canonical high/soft state in between. 
This is inferred even more clearly from the observed 
variability of Holmberg IX X1. A series of {\it ASCA} observations 
showed it in a hard power-law state (average $\Gamma \approx 1.6$: 
Ezoe, Iyomoto \& Makishima 2001) with $0.3$--$10$ keV luminosities 
varying between $\approx 0.6$--$2.0 \times 10^{40}$ erg s$^{-1}$; 
similarly, a 2002 {\it XMM-Newton} observations showed 
a hard power-law ($\Gamma \approx 1.8$) plus soft excess, with  
$L_{0.3-10} \approx 1.1 \times 10^{40}$ erg s$^{-1}$ (Miller, Fabian \& Miller 2004).
More recent {\it Swift} observations (Kaaret \& Feng 2009) 
show it in a hard state $\Gamma \approx 1.8$, variable by 
a factor of 7, with a peak luminosity $L_{0.3-10} \approx 2.8 \times 10^{40}$ erg s$^{-1}$.
But other {\it ASCA} observations in 1999 showed it in a curved 
or slim-disk state at $\approx 2.0 \times 10^{40}$ erg s$^{-1}$ (La Parola et al. 2001); 
and the same curved spectrum and luminosity was found in the 2001 {\it XMM-Newton} 
observations (Tsunoda et al.~2006).
Finally, to complicate things even further, we need to explain 
hard power-law ULXs that remained in that state despite luminosity variations 
of an order of magnitude: for example, NGC\,1365 X1 and X2 (Table 1).

\subsection{Evidence for a high/hard state}

It was suggested (Gladstone et al.~2009) that the hard state 
of ULXs is part of a continuous sequence of spectral states, 
all dominated by Comptonized emission, with decreasing electron 
temperature and increasing optical depth with accretion rate. 
In that sequence, the harder 
sources have accretion rates slightly above the maximum limit of 
the high/soft state, while ULXs with a steeper spectrum and high-energy 
downturn have even higher accretion rates and may be dominated 
by super-Eddington outflows. In this scenario, ULXs are not 
seen in the canonical high/soft state because their accretion rates 
are always above its threshold.
A minor difficulty of this linear sequence is that some variable ULXs 
have been observed in a hard power-law state across a large range of luminosities 
from $\sim$ a few $10^{39}$ erg s$^{-1}$ to $\sim$ a few $10^{40}$ erg s$^{-1}$.
If the hard ULX state always indicates a luminosity above the canonical 
high/soft state and below the curved soft state, some ULXs would have  
peak luminosities $\approx 10 L_{\rm Edd}$,  
which require accretion rates $\ga 1000$ Eddington. 
Another possible difficulty of the sequence proposed by 
Gladstone et al.~(2009) is that the observed fraction of hard-power-law ULXs 
is in fact greater within the most luminous group, 
above $10^{40}$ erg s$^{-1}$ (Berghea et al.~2008).

Taking into account the observational constraints, 
I suggest that hard-power-law ULXs represent a {\it persistent 
high/hard state} not seen in Galactic BHs. 
This hard state is not limited to luminosities $< 0.1 L_{\rm Edd}$, 
nor to a narrow luminosity range above the high/soft state. 
Instead, it may overlap in luminosity 
with all those states. At low accretion rates, it may coincide with 
the canonical low/hard state of Galactic BH transients.
But unlike those sources, I speculate that it may persist steadily up 
to much higher accretion rates, near or above the Eddington limit, 
at which point a standard disk would no longer be stable 
and a canonical hard/soft transition would not be possible anyway. 
Above those rates, some sources may switch directly to some varieties 
of warm corona or slim disk state, others may stay in a hard power-law state.
In other words, I speculate that BHs do not necessarily have to 
go through a disk-dominated, radio-quiet state when they 
vary between $\approx 0.01 L_{\rm Edd}$ and $\approx L_{\rm Edd}$.

\subsection{Suppressing the hard/soft switch?}

The relative balance of (harder) power-law and (softer) thermal disk components 
in the X-ray spectra of Galactic BHs may depend on various physical processes.
In one scenario, it depends on the relative contribution of the sub-Keplerian 
and Keplerian components of the accretion flow: the low-angular-momentum component 
feeds the hot scattering region, and the high-angular-momentum component 
feeds the thermal disk (Chakrabarti \& Titarchuk 1995; Smith, Heindl \& Swank 2002). 
In another scenario, it depends on the balance between evaporation 
and condensation at the disk/corona interface: at lower accretion rates 
the disk evaporates into the hot medium, and at higher accretion rates 
the hot corona or ADAF recondenses onto the disk (Meyer, Liu \& Meyer-Hofmeister 2000).

Observationally, hard-to-soft transitions in Galactic BH transients occur 
at luminosities $L \sim 0.02$--$0.1 L_{\rm Edd}$, at different luminosities 
for different sources, but also sometimes for the same source in different 
outbursts (Maccarone 2003; Yu \& Yan 2009; Qiao \& Liu 2009). 
This suggests that the accretion rate at which the geometrically 
thick, hot medium collapses onto the disk depends more on the outer boundary 
and internal conditions of the inflow than on the mass and spin of the BH.

Let us consider what physical parameters may determine whether 
the hot scattering region collapses at $\approx 0.03 L_{\rm Edd}$ 
or persists in a steady state up to $\ga L_{\rm Edd}$.
Most Galactic BH transients are fed by low-mass donor stars; 
instead, most ULXs are fed by an OB star. If the Comptonizing region 
is created by sub-Keplerian inflows, one might speculate that 
winds from an OB donor can efficiently replenish the hot structure.
However, other observed properties are not consistent with this scenario.
The accretion rates implied by ULX luminosities  
($\ga$ a few $10^{-6} M_{\odot}$ yr$^{-1}$) require  
mass transfer via Roche lobe overflow rather than wind accretion.
From that point of view, ULXs are more similar to low-mass X-ray binaries, 
with a large irradiated disk, than to wind-accreting high-mass X-ray binaries.
Also, many ULXs are located in moderately old stellar 
environments ($\sim 20$ Myr), suggesting B star rather that 
O star donors: this downplays the role of stellar winds.
And the two prototypical BH X-ray binaries with high-mass donors, 
Cyg X-1 and LMC X-1, do show transitions to the high/soft state; 
in fact, LMC X-1 is persistently in the high/soft state 
at an X-ray luminosity $\sim 0.1 L_{\rm Edd}$ (Yao, Wang \& Zhang 2005).

An alternative scenario is that the high/hard state of ULXs 
is due to a higher evaporation rate of the disk into the corona.
The corona/ADAF recondenses into a disk when the accretion rate 
exceeds the maximum evaporation rate. The latter strongly 
depends on the viscosity parameter in the disk (Qiao \& Liu 2009), 
via the competing effects of viscous heating and radial energy advection. 
For a Shakura-Sunyaev $\alpha$ parameter $\approx 0.1$--$0.9$, 
it was estimated that the hard/soft transition luminosity 
$\approx 0.38 \alpha^{2.34} L_{\rm Edd}$ (Qiao \& Liu 2009).
In this scenario, a high/hard state would require an effective 
viscosity parameter $\alpha \ga 1$ (perhaps dominated by magnetic viscosity).

Other physical parameters have been suggested 
to affect the threshold for the hard/soft transition 
and the relative size of the Comptonizing medium.
Magnetic fields may have a (comparatively weaker) effect 
on the evaporation rate via the heat conduction coefficient (Qian, Liu \& Wu 2007), 
and on the strength of the power-law emission, because synchrotron cooling 
in the corona competes against Compton cooling. 
The location of the (truncated) inner disk radius 
before the onset of an outburst was suggested to affect 
the transition luminosity (Zdziarski et al. 2004); 
and the hardness of the irradiating X-ray photons (Liu, Meyer \& Meyer-Hofmeister 2005). 
The rate of increase of the accretion rate and of the luminosity 
($dL/dt$) at the start of an outburst was shown to be correlated 
to the maximum luminosity in the hard state before the soft transition 
(Yu \& Yan 2009). But the last two effects have been studied 
in relation to transient sources; it is not clear why and how they 
could apply to ULXs in a persistently luminous and hard state. 
Another observational relation was shown to exist in transient 
stellar-mass BH and neutron star X-ray binaries, between the hard/soft 
transition luminosity (in Eddington units) and the duration 
of the quiescent phase before the outburst, which, in turn, is proportional 
to the mass stored in the disk (Yu, van der Klis \& Fender 2004; Yu et al.~2007).
It is not clear why this should be the case, and whether this relation 
extends to ULXs. But it is likely that ULXs have larger and more massive 
disks than Galactic BHs, if they have a Roche-lobe-filling OB donor star 
and BH masses $\approx 30$--$100 M_{\odot}$.

\section{ULX jets}

In Galactic BHs, hard/soft X-ray spectral transitions are associated 
with two other important changes. One is that the fractional broad-band 
variability decreases as the source becomes softer: rapid variability 
is strongly correlated with the Comptonized fraction of the luminosity 
(see Belloni 2010a for a review; and Belloni 2010b, in these proceedings).
The other change is that the steady jet (typical feature of the low/hard state) 
is suppressed (Fender et al.~2004), as the geometrically thick, hotter medium collapses 
onto the disk or is ejected. A flaring jet with intermittent blob ejections 
may, however, exist in the very high state.

If ULXs never switch to the high/soft state and never lose their dominant 
Comptonizing medium, can they also keep a steady jet up to the Eddington 
accretion rate? We do not know the answer to this question, yet.
However, we have recently found (Pakull, Soria \& Motch~2010; Soria et al.~2010)
a microquasar in NGC\,7793 with a pair of collimated jets, FRII-like radio lobes, 
and a bright, large radio nebula. We estimated a long-term-average jet power 
$\sim 10^{40}$ erg s$^{-1}$. The source is currently X-ray dim, but its mechanical 
power shows that steady collimated jets can persist even at accretion rates 
$\ga$ Eddington. 
Two other sources (in this case, X-ray luminous) have synchrotron radio nebulae 
suggesting a mechanical power $> 10^{39}$ erg s$^{-1}$: Holmberg II X1 
(Miller, Mushotzky \& Neff 2005) and NGC\,5408 X1 (Lang et al.~2007; 
Soria et al.~2006). But for those two ULXs, the presence of collimated jets 
has not yet been proven.
It is also possible that some ULXs have a jet and some do not; perhaps
this is the reason why some ULXs have been found to have much higher 
short-term variability than others, even though their X-ray spectra 
appear similar (Heil, Vaughan \& Roberts 2009).

\section{Comparison with AGN}

The low/hard versus high/soft state classification is often applied also 
to AGN (Jester 2005). I have argued that such scheme may not include or represent 
well at least one class of accreting BHs (ULXs). So, we need to check 
whether AGN show evidence of clear hard/soft transitions and at what 
luminosity range. 
Combining optical, UV and X-ray data, K\"{o}rding, Jester \& Fender (2006) showed 
that the AGN population is distributed in a similar way to Galactic BHs 
in a disk-fraction versus luminosity diagram ({\it i.e.}, measuring the relative 
contribution of the Comptonized component as a function of total luminosity).
However, it is not clear whether AGN undergo 
sharp hard/soft spectral transitions with a rapid collapse of the Comptonizing 
medium, or remain in intermediate states (see the discussion about the limited 
spectral evolution of Ark 564 in Belloni 2010a). 
Similarly to Galactic BHs, the average spectral index of the X-ray power-law component 
in AGN steepens at higher Eddington ratios (Shemmer et al.~2006). The relative fraction 
of power-law (coronal) emission was estimated at $\approx 45\%$ for a sample of AGN with 
Eddington ratio $0.0012 < L/L_{\rm Edd} < 0.032$, and $\approx 11\%$ for a sample with 
$0.27 < L/L_{\rm Edd} < 2.7$ (Vasudevan \& Fabian 2007), again suggesting a low/hard 
versus high/soft dichotomy. There is not enough information yet on what happens 
more specifically at $L/L_{\rm Edd} \ga 1$: whether the power-law/disk ratio increases again, 
and whether there is an AGN equivalent (most likely, in the radio-loud quasar class) 
of hard power-law-dominated ULXs. Narrow-Line Seyfert 1s may be emitting 
near or above Eddington, but they have a soft spectrum, similar to ULXs with 
a soft excess and a steep power law.

As far as the jet properties are concerned, there is some evidence 
(Maccarone, Gallo \& Fender 2003; Sikora, Stawarz \& Lasota 2007) 
of a three-state scenario, with FRI galaxies representing the radio-loud 
low/hard state, radio-quiet quasars representing the high/soft state, 
and radio-loud quasars and FRII galaxies representing the very high state.
In this scheme, ULXs would correspond to the last class.
It is not clear whether there is a sharp switch between radio states, 
and whether all AGN have to go through a radio-quiet 
gap in their evolution when they reach $L \sim 0.1 L_{\rm Edd}$, 
or can instead evolve directly between the FRI and FRII states.

\section{Conclusions}

I have argued that we face difficulties when we try to assign ULXs to 
one or more of the canonical accretion states, well defined in Galactic BHs.
Most ULXs with a soft excess, high-energy downturn and/or curved spectrum 
can be successfully explained with a variety of models (Comptonization 
in a warm corona, slim disk, outflows), representing an extension of 
the canonical very-high-state. But there is a significant fraction of ULXs 
with a hard power-law spectrum (including some of the most luminous ones), 
whose place in this scheme (or along a mass accretion rate sequence) 
remains unclear. In some cases, the hard power-law spectrum 
is preserved over luminosity variations of an order of magnitude.
Interpreting it as a low/hard state with an upper limit of $L \sim 0.1 L_{\rm Edd}$ 
would give implausibly high BH masses. More likely, this is 
a high/hard state, near or above Eddington, not known as 
a steady state in Galactic BHs. 

I speculated that some BHs may persist in a hard state 
over a large range of low and high accretion rates, 
and evolve directly to super-Eddington states, without going 
through a radio-quiet gap dominated by a standard disk with no corona.
If the high/soft state is suppressed in at least a subclass of ULXs, 
I suggested that they may also keep a steady jet up to Eddington 
accretion rates.

As for the ULXs with a curved spectrum, or an energy downturn 
at $\sim 5$ keV, it is now the general consensus that several 
models based on Comptonization in a warm corona plus outflows, as well as 
slim disk models, are consistent with the observations, 
and give a plausible BH mass range $\approx 30$--$100 M_{\odot}$ with 
luminosities $\la 3 L_{\rm Edd}$.
I proposed that bulk-motion Comptonization in the radiatively driven 
outflow from the inner disk may also be at work, and would produce  
a similar spectral downturn at $\sim 5$ keV.

\acknowledgements
I thank H. Feng, C. Motch, M. W. Pakull, T. P. Roberts, D. A. Swartz, D. J. Walton  
for their comments and suggestions. 


\end{document}